\documentclass[12pt,preprint]{aastex}
\slugcomment{Accepted for publication to New Astronomy}

\shorttitle{Micro-Variability of RQQSOs}
\shortauthors{Gupta \& Yuan}

\begin{document}


%
\title{Quasi-Simultaneous Two Band Optical Micro-Variability \\
    of Luminous Radio-Quiet QSOs}


\author{Alok C. Gupta\altaffilmark{1,2} and Weimin Yuan\altaffilmark{2}}
\altaffiltext{1}{Aryabhatta Research Institute of Observational Sciences (ARIES),
Manora Peak, \\
\hspace*{0.22in} Nainital $-$ 263129, India.}
\altaffiltext{2}{National Astronomical Observatories/Yunnan Observatory, Chinese
Academy of \\ 
\hspace*{0.22in} Sciences (CAS), P.O. Box 110, Kunming, Yunnan 650011, China.}

\email{acgupta30@gmail.com, wmy@ynao.ac.cn \\
Phone No. +91 9936683176, Fax No. +91 5942 233439}
%
%

%
%

\begin{abstract}
We report the first results of quasi-simultaneous two passband optical monitoring of 
six quasi-stellar objects to search for micro-variability. We carried out photometric 
monitoring of these sources in an alternating sequence of R and V passbands, for five 
radio-quiet quasi-stellar objects (RQQSOs), 0748$+$291, 0824$+$098, 0832$+$251, 
1101$+$319, 1225$+$317 and one radio-loud quasi-stellar object (RLQSO), 1410$+$429. 
No micro-variability was detected in any of the RQQSOs, but convincing micro-variability 
was detected in the RLQSO on two successive nights it was observed. Using the 
compiled data of optical micro-variability of RQQSOs till date, we got the duty cycle
for micro-variability in RQQSOs is $\sim$ 10\%. The present investigation indicates 
that micro-variability is not a persistent property of RQQSOs but an occasional
incident. 

\end{abstract}





\section{Introduction}

There are two major classes of quasars. The majority of these $\sim$ 85 $-$ 90 \% belong to 
the radio-quiet class and are thus known as radio-quiet quasars (RQQSOs) and the remaining 
$\sim$ 10 $-$ 15 \% are in the radio-loud class and are known as radio-loud quasars (RLQSOs).
RQQSOs and RLQSOs show, similar optical characteristics, but in RQQSOs the ratio of radio (at 
frequency $\nu =$ 5 GHz) to optical flux densities (at wavelength $\lambda =$ 4400$\AA$), 
R $\leq$ 10 (Kellerman et al. 1989), and the radio to X-ray fluxes are 1 to several orders of 
magnitude lower than those of the RLQSOs (Terashima \& Wilson 2003). 

Flux variability is a common property of active galactic nuclei (AGNs) and a small subset 
of radio-loud AGNs show variability on diverse time scales ranging from a few minutes to 
several years at almost all wavelengths of the EM spectrum, with the emission being strongly 
polarized. Such AGNs are called blazars, and their radiation at all wavelengths is predominantly 
nonthermal. Significant variability in brightness over a few minutes to several hours (less than 
a day) is commonly known as micro-variability, intra-night variability or intra-day variability. 
Optical micro-variability in blazars have been reported on several occasions in the last 
two decades using CCD detectors, and is now a well established property of blazars. The first 
pioneering papers in the field of blazar optical micro-variability using CCD detectors include: 
Miller, Carini \& Goodrich (1989); Carini et al. (1990, 1991, 1992); Carini \& Miller (1992); 
Heidt \& Wagner (1996). It is believed that the micro-variability in blazars is a consequence 
of relativistic beaming by jets (Bregmann 1991). The beaming can amplify intrinsic variations 
which may or may not originate within the jet. It is also well accepted that many other radio-loud 
AGNs (non blazars) also exhibit micro-variability (Jang \& Miller 1995, 1997; de Diego et al. 1998; 
Romero et al. 1999; Stalin et al. 2004, 2005). 

After about one and a half decades of work, detections of micro-variability in radio-quiet AGNs
have been elusive and little is known about their micro-variability. In a recent paper, 
Carini et al. (2007) compiled the micro-variability results for 117 radio-quiet AGNs to date, 
and in most of the cases the detected micro-variability is not convincing. Clear detection of 
micro-variability in a few radio-quiet QSOs have been reported in some recent papers 
(Gopal-Krishna et al. 2003; Stalin et al. 2004, 2005; Gupta \& Joshi 2005). 
Gupta \& Joshi (2005) compiled the micro-variability data for different classes of AGNs and have 
done statistical analysis of micro-variability behavior by dividing
the sample into 3 classes viz. radio-loud AGNs (blazars), radio-loud AGNs (non blazars) and 
radio-quiet AGNs. They found, generally $\approx$ 10\% and 35 $-$ 40\% radio-quiet AGNs and 
radio-loud AGNs (non blazars) respectively show micro-variability. For any blazar, if 
observed continuously for less than 6 hours and more than 6 hours, the chances of detection
of micro-variability are $\approx$ 60 $-$ 65\% and 80 $-$ 85\% respectively. They also found 
that the maximum amplitudes of variations are about 10\%, 50\% and 100\% for radio-quiet AGNs, 
radio-loud AGNs (non blazars) and blazars respectively. 

So far, for optical micro-variability studies of RQQSOs and RLQSOs, several groups have observed 
sources continuously in only one optical band (Gupta \& Joshi 2005; Carini et al. 2007 and references 
therein), and clear detections in a few cases of micro-variability in RQQSOs suggest micro-variation 
may be caused by a faint jet (Gopal-Krishna et al. 2000; Falcke et al. 1996b; Ulvestad et al. 2005) 
or optical flares above an accretion disk and/or accretion disk instabilities (Wiita et al. 1992). 
In the present work, a sample of five RQQSOs and one RLQSO are considered for study. These five 
RQQSOs have shown optical micro-variability on some occasions in their earlier observations. However, 
to investigate whether micro-variability is a persistent property of RQQSOs or it is only occasional 
events, we monitored these sources again. We also included a new radio-loud QSO to search for optical 
micro-variability. Here we are reporting the quasi-simultaneous continuous observations of these 
sources in two optical bands (V and R) in an alternating sequence for these sources for the first time. 
In addition to looking for micro-variability in two optical bands, our observations also give additional 
information of micro-variation in V$-$R color. 

Section 2 presents our sample selection criterion; in Section 3, we report the observations 
and data analysis technique; in Section 4, the results of the present work are presented and
in Section 5 our conclusions are given.   
 
\section{Sample Selection Criterion}

In the present study, we have selected five RQQSOs and one RLQSO from the recent catalog of 
active galactic nuclei and quasars of Veron-Cetty \& Veron (2006). The detailed information 
and observation log of the QSOs are given in Table 1. Using the standard cosmological 
model, we have used the well known relation by Weinberg (1972) for calculating the absolute 
magnitude of the QSOs (M$_{V}$),
\begin{eqnarray}
m_{v} - M_{v} = 25 - 5 \rm{log}H_{0}  \ \rm(km  \ s^{-1} \ Mpc^{-1}) + 5 \rm{logcz} \ \rm(km \ s^{-1}) 
+ 1.086 (1-q_{0}) z
\end{eqnarray}
where m$_{v}$, $z$ are, respectively, the apparent magnitude and redshift of the QSOs, and 
$c$ the speed of light. We have used the Hubble constant H$_{0}$ = 70 km s$^{-1}$ Mpc$^{-1}$ and 
q$_{0}$ = 0.5. 

The host galaxy is expected to contribute less than 10\% to the total flux of the luminous QSOs. 
The host galaxy is also expected to be encompassed within the aperture radius $\sim$ 6 arcsec used 
for photometry. For a lower ratio of AGNs to galactic light, false indications of variability produced 
by seeing variations that include different amounts of host galactic light within the photometric 
aperture become very probable (Cellone et al. 2000). Carini et al. (1991) investigated whether 
a conspicuous galaxy component produces variations, due to variation in atmospheric seeing or 
transparency, which are not intrinsic to the source. They showed that, even for sources with 
significant underlying galaxy components, any spurious variations introduced by fluctuations in 
atmospheric seeing or transparency are typically smaller than the observational uncertainties.
To further reduce this effect, we have selected sources which are optically bright (brighter than 
M$_{V} < -$24.9 mag) (except one RQQSO namely Ton 52 which has M$_{V} = -$24.3 mag), so that 
the fluctuations due to the underlying galaxy are minimal; however the modest optical luminosity 
(M$_{V} \approx -$24.9 mag) lies close to the critical value below which the sources are classified 
as Seyfert galaxies (Miller et al. 1990). 

\section{Observations and Data Reductions}

The photometric observations of the five radio-quiet QSOs and one radio-loud QSO were 
carried out in the V and R passbands of the optical filter system of Bessell (1990) using 
a Loral Lick 3 CCD detector (2048 pixels $\times$ 2048 pixels) mounted 
at the f/9 Cassegrain focus of the 2.16 meter R-C system optical telescope at the National 
Astronomical Observatories, Xinglong Station, in China. Observations were carried out 
using BFOSC (Bao Faint Object Spectrograph and Camera) in an alternating sequence of V and R 
passbands. The pixel size of the CCD detector is 15 $\mu$m $\times$ 15 $\mu$m and each pixel of the 
CCD projected on the sky corresponds to 0.305 arcsec in both dimensions. The entire CCD 
chip covers $\sim$ 10.41 $\times$ 10.41 arcmin$^{2}$ of the sky. Read out noise and gain of 
the CCD detector were 1.67 electrons and 4 electrons/ADU, respectively. Throughout the observing 
run, the typical seeing was $\sim$ 2.0 arcsec ranging from 1.5 to 2.5 arcsec. Several bias frames 
were taken intermittently in each observing night and twilight sky flats were taken in V and R 
passbands. The observation log is given in Table 1.

We constructed median bias, and median flat field images in the V and R passbands for each night 
which were used for bias and flat field corrections. Image processing or pre processing (bias 
subtraction, flat-fielding and cosmic rays removal) were done using standard routines in 
IRAF\footnote{IRAF is distributed by the National Optical Astronomy Observatories, which are 
operated by the Association of Universities for Research in Astronomy, Inc., under cooperative 
agreement with the National Science Foundation.} (Image Reduction and Analysis Facility) 
software. Photometric reduction (instrumental magnitude of the QSOs and local comparison stars 
in the QSOs fields) of the data were performed by aperture photometric technique using DAOPHOT II 
(Dominian Astronomical Observatory Photometry) software (Stetson 1987). Aperture photometry was 
carried out with 4 concentric apertures of radii of $\approx$ 1, 2, 3 and 4 times of the FWHM of 
stars in the image frames. The data reduced with different aperture radii were found to be in good 
agreement. However, it was noticed that the best signal to noise ratio (S/N) was obtained with an 
aperture radius of $\approx$ 3 $\times$ the typical FWHM. In each QSO field, we selected 4 comparison 
stars and finally used the 2 best non-variable stars for analysis purposes (constructing differential 
instrumental magnitude light curves). The coordinates and B and R magnitude of the 4 comparison stars 
in each QSO field are given in Table 2, which are taken from the USNO (United States Naval Observatory) 
catalog (Monet et al. 2003). The coordinates and V magnitude of comparison stars were taken from STScI 
(Space Telescope Science Institute) GSC 2.2 (Guide Star Catalog Version 2.2), if stars were not available 
in the USNO catalog. The positional accuracy in GSC 2.2 is 0.3 arcsec.  

In the present work, in general, we have got the photometric error in each data point is less 
than 0.01 magnitude ($\sim$ better than 1\%). If we take 3$\sigma$ detection of micro-variabilty as 
genuine micro-variability then any variation in magnitude or color is more than 0.03 magnitude 
($\sim$ more than 3\%) should be clearly visible. Observations presented here have maximum exposure 
time of 250 seconds in V band followed by 150 seconds in R band. After including the readout time 
of the CCD detector, we repeat our observations in V or R band maximum after $\sim$ 10 minutes. So, 
the data presented here is sensitive for the minimum time scale of 10 minutes and any detected 
variability with time scale more than 10 minutes should be clearly visible.   

\section{Results}

Differential Light Curves (DLCs) of the five RQQSOs and the RLQSO, are plotted in Figs. 1$-$6. From 
bottom to top, the panels show the differential instrumental magnitudes of the QSO and two comparison 
stars in the R band, V band and V-R color. In the figures Q and S represent the QSO and the comparison 
star, respectively. For generating V$-$R light curves, we have taken an average time for each set of 
alternate image frames and the R band differential magnitude is subtracted from the corresponding V 
band differential magnitude. We use offsets to plot the lights curves for clarity in the figures.

\subsection{Micro-Variability and Variability Amplitude}

Using the aperture photometry of the QSO and two comparison stars in the QSO field, 
we determined the differential instrumental magnitudes V, R and V$-$R; of the QSO $-$ comparison 
star A, QSO $-$ comparison star B and comparison star A $-$ comparison star B. We determined the
respective observational scatter from; QSO $-$ comparison star A $\sigma$ (QSO - Star A), QSO 
$-$ comparison star B $\sigma$ (QSO - Star B) and comparison star A $-$ comparison star B
$\sigma$ (Star A - Star B). The variability of the target QSO is quantified by the variability 
parameter, C, introduced by Romero et al. (1999); this variability parameter is expressed as
the average of C$_{1}$ and C$_{2}$

\begin{eqnarray}
C_{1} = \frac {\sigma (QSO - Star A)}{\sigma (Star A - Star B)} \hspace*{0.2in} \rm{and} \hspace*{0.2in}
C_{2} = \frac {\sigma (QSO - Star B)}{\sigma (Star A - Star B)}
\end{eqnarray}
If C $>$ 2.57, the confidence level of variability is 99\%, and we follow most previous authors in
adopting this conservative criterion.. 
The value of C by using both comparison stars for all of the five radio-quiet QSOs and one radio-loud QSO 
for different observing nights are reported in table 3. 

We use the intra-day variability amplitude defined by Heidt \& Wagner (1996)
\begin{eqnarray}
A = 100 \times \sqrt {(A_{max} - A_{min})^{2} - 2\sigma^{2}} \hspace*{0.1in}\%
\end{eqnarray}

where A$_{max}$ and A$_{min}$ are the maximum and minimum in the differential light curve and
$\sigma$ the measurement errors. 
The measured amplitudes are reported in Table 3.

\subsection{Notes on Individual Sources}

\noindent
{\bf 0748$+$291 (QJ 0751$+$2991)}

This RQQSO, reported as the brightest new QSO in the first bright QSO survey (Gregg et al.
1996), has been monitored to search for micro-variability in optical bands on several occasions 
(Gopal-Krishna et al. 2000; Stalin et al. 2004; Gupta \& Joshi 2005). In one night of observations, 
Gopal-Krishna et al. (2000) reported detection of spikes (brightness excursions of only a single point). 
Stalin et al. (2004) did not find any micro-variability in the source in their six nights of observations 
spread over more than three years, but long term variations are clearly seen in their observations. Gupta 
\& Joshi (2005) reported the clear detection of micro-variation in the source of their monitoring of the 
source continuously for 8 hours in the V-passband in one night.

We observed the source during the night of March 12, 2007. DLCs are plotted in Fig. 1. By using 
our micro-variability detection test given in equation (2), we found the values of C in V, R and (V-R) 
are 1.0, 1.9 and 1.2 respectively. These values of C show that this RQQSO did not show any genuine 
micro-variation in V, R and V-R observations. 

\noindent
{\bf 0824$+$098 (1WGA J0827.6$+$0942)} 

This RQQSO has been monitored to search for micro-variability in the optical R band on two occasions 
(Gopal-Krishna et al. 2000; Stalin et al. 2005). Stalin et al. (2005) found 2.2\% micro-variation 
in the source in 8.2 hours of continuous monitoring over one night (December 27, 1998). In 3.3 hours 
of observations in one night (February 15, 1999), Gopal-Krishna et al. (2000) reported detection of 
one spike.

We observed the source during the night of March 11, 2007. DLCs are plotted in Fig. 2. We found that 
the values of C in V, R and (V-R) are 1.2, 1.7 and 1.2 respectively. These values of C show that this 
RQQSO did not show any genuine micro-variation in V, R and V$-$R. At the end of the light curve in 
the V band, the source appeared to become $\sim$ 0.05 mag fainter and came back to its the normal 
magnitude in $\sim$ 20 minutes. This nominal variation in the V band is also transmitted in the 
V $-$ R color and made a change of $\sim$ 0.05 mag in the same duration of observations. Confirmation 
of such events require further monitoring of the source for longer durations with similar or better S/N.     

\noindent
{\bf 0832$+$251 (PG 0832$+$251)}

This RQQSO was observed in 3 nights in less than one year in a search for micro-variability 
(Stalin et al. 2005). Micro-variability was detected in the source on one night.  

We observed the source during the night of March 10, 2007. DLCs are plotted in Fig. 3. We found that 
the values of C for V, R and V-R are 1.1, 1.0 and 1.1 respectively. These value of C show that this 
RQQSO did not show any genuine micro-variation in V, R and V$-$R. 

\noindent
{\bf 1101$+$319 (Ton 52)}

This RQQSO was observed in 5 nights spread over about 2 years in a search for micro-variability 
(Stalin et al. 2004; Gupta \& Joshi 2005). Micro-variability was detected in the source on one
night and long term variation was also noticed (Stalin et al. 2004).  

We observed the source during the night of March 11, 2007. DLCs are plotted in Fig. 4. In the
DLCs, the last data point in the V band has a comparatively large error bar. In our variability 
detection test given by equation (2), we omitted this data point and found that the values of C for 
V, R and (V-R) are 1.3, 1.5 and 1.8 respectively. These values of C show that there was no genuine 
micro-variation in V, R and V-R for this RQQSO.    

\noindent
{\bf 1225$+$317 (b2 1225$+$317)}

This source has only been monitored for one night in the V passband for 6.2 continuous hours
in a search for micro-variability (Gupta \& Joshi 2005). There was a possible detection of 
micro-variability in the source.  

We observed the source during the night of March 10, 2007. DLCs are plotted in Fig. 5. We found the 
values of C in V, R and V-R are 1.4, 1.8 and 2.3 respectively. The value of C shows that this RQQSO did 
not exhibit any micro-variation in the V, R and V-R color. 

\noindent
{\bf 1410$+$429 (RXS J14119$+$4239)}

This source is the only radio-loud QSO studied in the present work. So far, this source has not
been studied to search for optical micro-variability.

We observed the source continuously (UT 19.12 to 21.59 hour) and (UT 18.12 to 21.52 hour) during 
the nights of March 11 and March 12, 2007 respectively. We obtained the DLCs for both nights of 
observations and are plotted in Fig. 6. We found the values of C in V, R and V-R and they are 3.0, 
3.8 and 4.3 respectively for the March 11 observations. For the observations on March 12, the values 
of C in V, R and V-R are 2.8, 5.0 and 3.7 respectively. On March 11, one spike was detected in the 
R band observations which changed the V-R color of the RLQSO. In determination of the values of C, 
the spike point has been omitted. These values of C show that this RLQSO exhibits clear micro-variation 
in V, R and V-R on both nights of observation. For the March 11, 2007 observations the amplitudes of 
variability in V, R and V-R are 4.0\%, 3.5\% and 6.6\%, respectively. For the March 12, 2007 observations 
the amplitudes of variability in V, R and V-R are 3.9\%, 6.5\% and 13.7\%, respectively. The differential 
magnitude of the RLQSO shows large error bars because it is $\sim$ 2.0 magnitudes fainter than both 
comparison stars. 

\subsection{Duty Cycle of Micro-variability in RQQSOs}

Since 1993, there are several attempts to search for optical micro-variability in radio-quiet 
AGNs by various groups around the globe (e.g. Gupta \& Joshi 2005, Carini et al. 2007 and references 
therein). But there are only a few occasions on which micro-variability is detected (see the  
compiled optical micro-variability results in Table 3. of Carini et al. 2007). In last one and half 
decade observations were carried out in search for optical micro-variability were at non regular 
time interval, inconsistent observing time duration for different sources, different selection 
criterion and different data analysis methods of sources by different groups. Here we have taken 
the compiled data of RQQSOs (not Seyfert Galaxies) from the Table 3 of Carini et al. (2007) and 
data of the 5 RQQSOs from the present work to calculate the duty cycle of micro-variability in 
RQQSOs. Here we defined the duty cycle without weighting the number of hours sources have been 
observed. If $n$ denotes the total number of occasions on which the optical micro-variability 
have been detected, and $N$ denotes the total number of occasions on which the sources have been 
observed in search for micro-variability then the duty cycle DC is defined as,
\begin{eqnarray}
\rm{DC} = {\frac {n}{N}} \time 100\%
\end{eqnarray}
Till date, 70 RQQSOs are observed at 217 occasions in search for optical micro-variability. 
Optical micro-variability detected in only 17 RQQSOs on 21 occasions in 88 occasions 
they have been observed. While 53 RQQSOs have never shown optical micro-variability in their 
observations on 129 occasions. So, the duty cycle of optical micro-variability detection in 
RQQSOs is only $\sim$ 10\%.

\section{Conclusions}

To investigate whether micro-variability is a persistent, or only occasional, property of RQQSOs,
we have presented new observations of five luminous radio-quiet QSOs and one radio-loud QSO 
in a search for quasi-simultaneous optical micro-variability in V and R passbands and V-R color for 
the first time. The RQQSOs studied here have shown optical micro-variability on some occasions of
their earlier observations. We found genuine micro-variations in one radio-loud QSO, (1410$+$429) in 
V, R and V-R on both nights for which the source was observed. Five radio-quiet QSOs, 0748$+$294, 
0824$+$098, 0832$+$251, 1101$+$319 and 1225$+$317 did not show any micro-variation in V, R and V-R. 
Our observations show that micro-variation in RQQSOs is rather rare. Non detection of 
micro-variability in any of the RQQSOs in our search is consistent with the other studies (see 
Table 3 of Carini et al. 2007). 

The generally accepted model for micro-variation in radio-loud AGNs is the shock-in-jet model (e.g. 
Blandford \& K$\ddot{o}$nigl 1979; Scheuer \& Readhead 1979; Marscher 1980; Hughes et al. 1985; 
Marscher 1992; Marscher \& Gear 1985; Valtaoja et al. 1988; Qian et al. 1991), in which the light is 
seen to fluctuate on time scales of a few minutes to an hour. Other models that can explain the 
micro-variation in any type of AGNs are optical flares, disturbances or hot spots on the accretion disk 
surrounding the black hole of the AGNs (e.g. Wiita et al. 1991, 1992; Chakrabarti \& Wiita 1993; 
Mangalam \& Wiita 1993). The micro-variation detected in V, R and V-R of the RLQSO 1410+429 can be 
explained by any of these models. 

After intensive searches of micro-variation in RQQSOs 
over the last one and half decades, the real cause of micro-variation is still not well known. Possible 
explanations involve either by shock-in-jet or accretion disk based models (Gupta \& Joshi 2005). 
It is believed that, in radio-quiet AGNs, due to severe inverse-Compton losses, most jets are probably 
quenched at the incipient stage (Wilson \& Colbert 1995; Blandford 2000). The weak and rarer evidence 
of optical micro-variation detection in earlier observations of RQQSOs can be explained with jet models 
that assume less Doppler boosting than the radio-loud QSOs and much less than the blazars (Gopal-Krishna 
et al. 2003), which is consistent with unified models of AGN. VLBA and XMM-Newton studies confirm weak 
jet emissions in radio and X-ray bands from a few radio-quiet QSOs on some occasions. Some evidence 
of weak radio jets in a small number of RQQSOs have been reported from deep VLA imaging and related 
studies (Miller et al. 1993; Kellermann et al. 1994; Falcke et al. 1996a,b). There have been some attempts 
at VLBA imaging at milliarcsecond resolution of the central engines of 12 radio-quiet QSOs by Blundell 
\& Beasley (1998). They reported 8 of these sources show strong evidence of a jet-producing central 
engine. In another recent paper, Ulvestad et al. (2005) have done deep VLBA imaging of 5 RQQSOs in which 
only one source shows a two-sided radio jet and the other 4 are unresolved. In radio monitoring campaigns 
with VLBA, relativistic jets were found in the RQQSO PG 1407$+$263 and in Seyfert 1 galaxy III Zw 2 
(Blundell et al. 2003; Brunthaler et al. 2005). Gallo (2006) in a recent paper has reported XMM-Newton 
observations of a RQQSO PG 1407$+$265 in which the X-ray variable emission apparently originates 
from a combination of jet and accretion disk processes, and where a relativistic X-ray jet only works 
intermittently. 

In a recent paper, Czerny et al. (2008) have done modeling of micro-variability of RQQSOs, using 
non-simultaneous X-ray and optical data of 10 RQQSOs which have shown optical micro-variability on some 
occasions. They have discussed that the three possible models for micro-variability in RQQSOs: 
(i) irradiation of an accretion disc by a variable X-ray flux, (ii) an accretion disc instability, 
(iii) the presence of weak blazar component, if jet emission is variable. They concluded that the 
blazar component model is the most promising model to explain the micro-variability in RQQSOs. 
Future simultaneous multi-color optical and X-ray monitoring observations of a sample of RQQSOs 
will give insight into the cause of micro-variation in RQQSOs. Micro-variability of RQQSOs 
is not yet well known and it needs a focused effort which we have plan to do in near future. 

\acknowledgments

We thank the anonymous referees for their helpful suggestions. We are grateful to Prof. P. J. Wiita 
for careful reading of the manuscript and his comments and suggestions. 
Dr. R. S. Pokorny is thankfully acknowledged for correcting grammatical mistakes. We are thankful to the 
2.16 meter telescope time allocation committee for allocation of observing time for the project. ACG is 
grateful for hospitality at National Astronomical Observatories, Xinglong Station, China during observing 
run and to Mr. Jia Junjun for helping in observations. We gratefully acknowledge the financial support from 
the National Natural Science Foundation of China (grant no. 10533050).

\clearpage

\begin{deluxetable}{ccccccccccc}
\tabletypesize{\scriptsize}
\rotate
\tablecaption{Complete log of V and R bands observations of five RQQSOs and one RLQSO$^{***}$. 
\label{tbl-1}}
\tablewidth{0pt}
\tablehead{
\colhead{IAU Name$^{*}$} & \colhead{Other Name} & \colhead{$\alpha_{2000.0}$} 
& \colhead{$\delta_{2000.0}$} & \colhead{z} & \colhead{V} & \colhead{M$_{V}$} 
& R$^{**}$ & \colhead{Date of Obs.} & \colhead{Data Points}          & \colhead{Data Points} \\
                         &                      &      
&                             &             &             &
&   &                        & \colhead{$\times$Exp. Time (V)}& \colhead{$\times$Exp. Time (R)} 
}
\startdata
0748$+$294 & Q J0751$+$2919 & 07 51 12.3 & $+$29 19 38 & 0.912 & 16.21 & $-$27.2 & 0.21
& 12. 03. 2007 & 44$\times$120 sec & 44$\times$80 sec \\
0824$+$098 & 1WGA J0827.6$+$0942 & 08 27 40.1 & $+$09 42 10.0 & 0.260 & 15.5 & $-$24.9 & 3.2
& 11. 03. 2007 & 40$\times$90 sec & 40$\times$60 sec    \\
0832$+$251 & PG 0832$+$251 & 08 35 35.9 & $+$24 59 41.0 & 0.331 & 16.1 & $-$24.9 & 1.26
& 10. 03. 2007 & 38$\times$120 sec & 39$\times$80 sec \\
1101$+$319 & Ton 52         & 11 04 07.0 & $+$31 41 11 & 0.440 & 17.30 & $-$24.3 & $<$0.39
& 11. 03. 2007 & 22$\times$150 sec & 22$\times$100 sec  \\
1225$+$317 & b2 1225$+$317  & 12 28 24.8 & $+$31 28 38 & 2.219 & 15.87 & $-$30.2 & ......
& 10. 03. 2007 & 44$\times$120 sec & 44$\times$80 sec  \\
1410$+$429 & RXS J14119$+$4239 & 14 11 59.7 & $+$42 39 50.0 & 0.888 & 17.37 & $-$26.0 & $>$155
& 11. 03. 2007 & 18$\times$150 sec & 18$\times$100 sec  \\
1410$+$429 & RXS J14119$+$4239 & 14 11 59.7 & $+$42 39 50.0 & 0.888 & 17.37 & $-$26.0 & $>$155
& 12. 03. 2007 & 20$\times$250 sec & 20$\times$150 sec   \\
\enddata


%
\tablenotetext{*} {Based on coordinates defined for 1950.0 epoch.} 
\tablenotetext{**} {The value of R for RQQSOs were taken from Carini et al. (2007).} 
\tablenotetext{***} {Other parameters of QSOs were taken from the Veron-Cetty \& Veron
(2006).} 

\end{deluxetable}

\clearpage

\begin{deluxetable}{cccccccl}
\tabletypesize{\scriptsize}
\tablecaption{Coordinates and apparent magnitudes of comparison stars in the field of
observed RQQSOs and RLQSO
\label{tbl-2}}
\tablewidth{0pt}
\tablehead{
\colhead{IAU Name} & \colhead{Star No.} & \colhead{$\alpha_{2000.0}$} 
& \colhead{$\delta_{2000.0}$} & \colhead{B (mag)} & \colhead{V (mag)} & \colhead{R (mag)} & \colhead{Remarks}
}
\startdata
0748$+$291 & 1 & 07 51 09.7 & $+$29 21 05.9 &      & 14.6$\pm$0.5 &      &    \\
           & 2 & 07 51 02.6 & $+$29 19 23.9 & 16.5 &              & 15.0 &    \\
           & 3 & 07 51 09.2 & $+$29 16 19.9 & 17.0 &              & 15.5 &    \\
           & 4 & 07 50 59.0 & $+$29 16 49.8 &      & 15.3$\pm$0.6 &      &    \\
0824$+$098 & 1 & 08 27 44.3 & $+$09 45 05.3 & 17.3 &              & 15.7 &    \\
           & 2 &            &               &      &              &      & Not detected in USNO and GSC 2.2 \\
           & 3 & 08 27 34.6 & $+$09 39 24.6 & 16.1 &              & 15.0 &    \\
           & 4 & 08 27 23.5 & $+$09 41 17.5 &      & 15.1$\pm$0.6 &      &    \\
0832$+$251 & 1 & 08 35 24.0 & $+$25 01 04.6 & 16.8 &              & 15.1 &    \\
           & 2 & 08 35 12.0 & $+$24 57 12.3 &      & 15.4$\pm$0.6 &      &    \\
           & 3 & 08 35 44.1 & $+$25 02 52.1 &      & 15.2$\pm$0.6 &      &    \\
           & 4 & 08 35 47.3 & $+$24 57 19.3 & 16.9 &              & 15.4 &    \\
1101$+$319 & 1 & 11 04 21.2 & $+$31 47 58.2 & 17.4 &              & 16.7 &    \\
           & 2 &            &               &      &              &      & Not detected in USNO and GSC 2.2 \\
           & 3 & 11 04 14.1 & $+$31 44 09.1 & 18.5 &              & 15.9 &    \\
           & 4 & 11 03 59.5 & $+$31 47 20.7 & 16.7 &              & 15.4 &    \\
1225$+$317 & 1 & 12 28 11.0 & $+$31 27 18.9 &      & 14.6$\pm$0.6 &      &    \\ 
           & 2 & 12 28 30.6 & $+$31 26 33.5 & 16.7 &              & 15.6 &    \\
           & 3 & 12 28 18.7 & $+$31 25 19.7 &      & 15.3$\pm$0.6 &      &    \\
           & 4 & 12 27 55.2 & $+$31 31 53.3 &      & 15.3$\pm$0.6 &      &    \\
1410$+$429 & 1 & 14 11 38.8 & $+$42 44 08.3 &      & 14.8$\pm$0.6 &      &    \\
           & 2 & 14 11 29.9 & $+$42 43 32.5 &      & 14.6$\pm$0.6 &      &    \\
           & 3 & 14 11 46.4 & $+$42 43 28.5 & 17.5 &              & 16.0 &    \\
           & 4 & 14 11 54.0 & $+$42 41 22.3 &      & 15.0$\pm$0.6 &      &    \\
\enddata


%

\end{deluxetable}

\clearpage

\begin{deluxetable}{cccccrrrcccl}
\tabletypesize{\scriptsize}
\rotate
\tablecaption{Results of micro-variability observations of RQQSOs and RLQSO$^{*}$. 
\label{tbl-3}}
\tablewidth{0pt}
\tablehead{
\colhead{Date} & \colhead{QSO} & \colhead{class} & \colhead{Band} & \colhead{N} & \colhead{Diff. mag}     & \colhead{Diff. mag} 
& \colhead{Diff. mag}         & \colhead{Variable} & \colhead{C} & \colhead{A\%} & \colhead{Remarks} \\
\colhead{dd.mm.yyyy} &        &         &                &             & \colhead{QSQ - S$_{A}$} & \colhead{QSQ - S$_{B}$}             
& \colhead{S$_{A}$ - S$_{B}$} &          &             &  &
}
\startdata
12.03.2007 & 0748$+$294 & RQQSO & V     & 33 & 0.926$\pm$0.009    & 0.232$\pm$0.009    & $-$0.694$\pm$0.009 & NV & 1.0 & &  \\ 
           &            &       & R     & 33 & 1.515$\pm$0.010    & 0.576$\pm$0.009    & $-$0.940$\pm$0.005 & NV & 1.9 & &  \\
           &            &       & V$-$R & 33 & $-$0.589$\pm$0.009 & $-$0.344$\pm$0.010 & 0.245$\pm$0.008    & NV & 1.2 & &  \\
11.03.2007 & 0824$+$098 & RQQSO & V     & 40 & 1.239$\pm$0.010    & 0.805$\pm$0.009    & $-$0.434$\pm$0.008 & NV & 1.2 & &  \\ 
           &            &       & R     & 40 & 0.765$\pm$0.010    & 0.546$\pm$0.010    & $-$0.219$\pm$0.006 & NV & 1.7 & &  \\
           &            &       & V$-$R & 40 & 0.474$\pm$0.012    & 0.259$\pm$0.012    & $-$0.216$\pm$0.010 & NV & 1.2 & &  \\
10.03.2007 & 0832$+$251 & RQQSO & V     & 38 & 0.296$\pm$0.005    & 0.492$\pm$0.006    &    0.197$\pm$0.005 & NV & 1.1 & &  \\ 
           &            &       & R     & 39 & 0.382$\pm$0.006    & 0.521$\pm$0.006    &    0.139$\pm$0.006 & NV & 1.0 & &  \\
           &            &       & V$-$R & 38 & $-$0.087$\pm$0.007 & $-$0.029$\pm$0.008 &    0.058$\pm$0.007 & NV & 1.1 & &  \\
11.03.2007 & 1101$+$319 & RQQSO & V     & 20 & $-$0.127$\pm$0.009 & 0.806$\pm$0.007    &    0.933$\pm$0.006 & NV & 1.3 & & Last V passband \\ 
           &            &       & R     & 21 & 0.424$\pm$0.007    & 0.907$\pm$0.008    &    0.483$\pm$0.005 & NV & 1.5 & & point omitted \\
           &            &       & V$-$R & 20 & $-$0.551$\pm$0.012 & $-$0.101$\pm$0.012 &    0.450$\pm$0.007 & NV & 1.8 & & in analysis  \\
10.03.2007 & 1225$+$317 & RQQSO & V     & 44 & 1.525$\pm$0.006    & 1.140$\pm$0.005    & $-$0.386$\pm$0.004 & NV & 1.4 & &  \\ 
           &            &       & R     & 44 & 1.555$\pm$0.007    & 1.219$\pm$0.007    & $-$0.336$\pm$0.004 & NV & 1.8 & &  \\
           &            &       & V$-$R & 44 & $-$0.029$\pm$0.009 & $-$0.079$\pm$0.009 & $-$0.050$\pm$0.004 & NV & 2.3 & &  \\
11.03.2007 & 1410$+$429 & RLQSO & V     & 18 & 2.028$\pm$0.009    & 2.122$\pm$0.009    &    0.094$\pm$0.003 & ~V & 3.0 & ~4.0 & one point spike in \\ 
           &            &       & R     & 17 & 2.231$\pm$0.012    & 2.554$\pm$0.011    &    0.323$\pm$0.003 & ~V & 3.8 & ~3.5 & R passband is removed \\
           &            &       & V$-$R & 17 & $-$0.203$\pm$0.017 & $-$0.433$\pm$0.017 & $-$0.229$\pm$0.004 & ~V & 4.3 & ~6.6 & in analysis \\
12.03.2007 & 1410$+$429 & RLQSO & V     & 20 & 1.998$\pm$0.010    & 2.110$\pm$0.012    &    0.111$\pm$0.004 & ~V & 2.8 & ~3.9 &  \\ 
           &            &       & R     & 20 & 2.111$\pm$0.021    & 2.547$\pm$0.019    &    0.336$\pm$0.004 & ~V & 5.0 & ~6.5 &  \\
           &            &       & V$-$R & 20 & $-$0.213$\pm$0.023 & $-$0.437$\pm$0.021 & $-$0.224$\pm$0.006 & ~V & 3.7 & 13.7 &  \\
\enddata


%
\tablenotetext{*} {V and NV in the Variable column represent variable and non variable respectively. N represent the
number of data points.}

\end{deluxetable}

\clearpage

\begin{figure}
\plotone{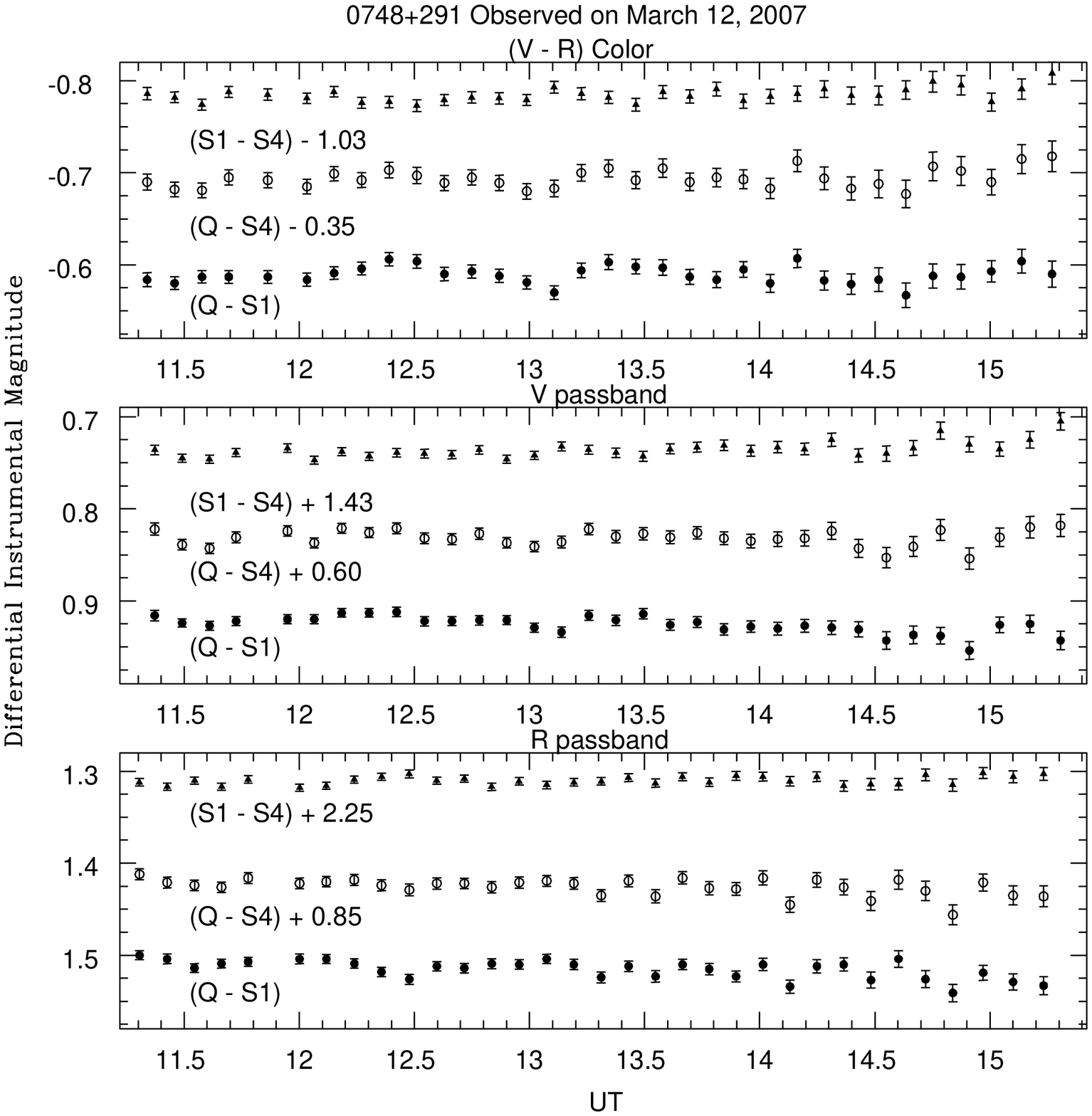}
\caption{The V, R and V$-$R light curves of 0748$+$294 on the night of March 12, 2007. 
In caption S and Q represent star and QSO respectively. For clarity, the DLCs are offseted 
by the amounts marked on the panels.}
\end{figure}

\begin{figure}
\plotone{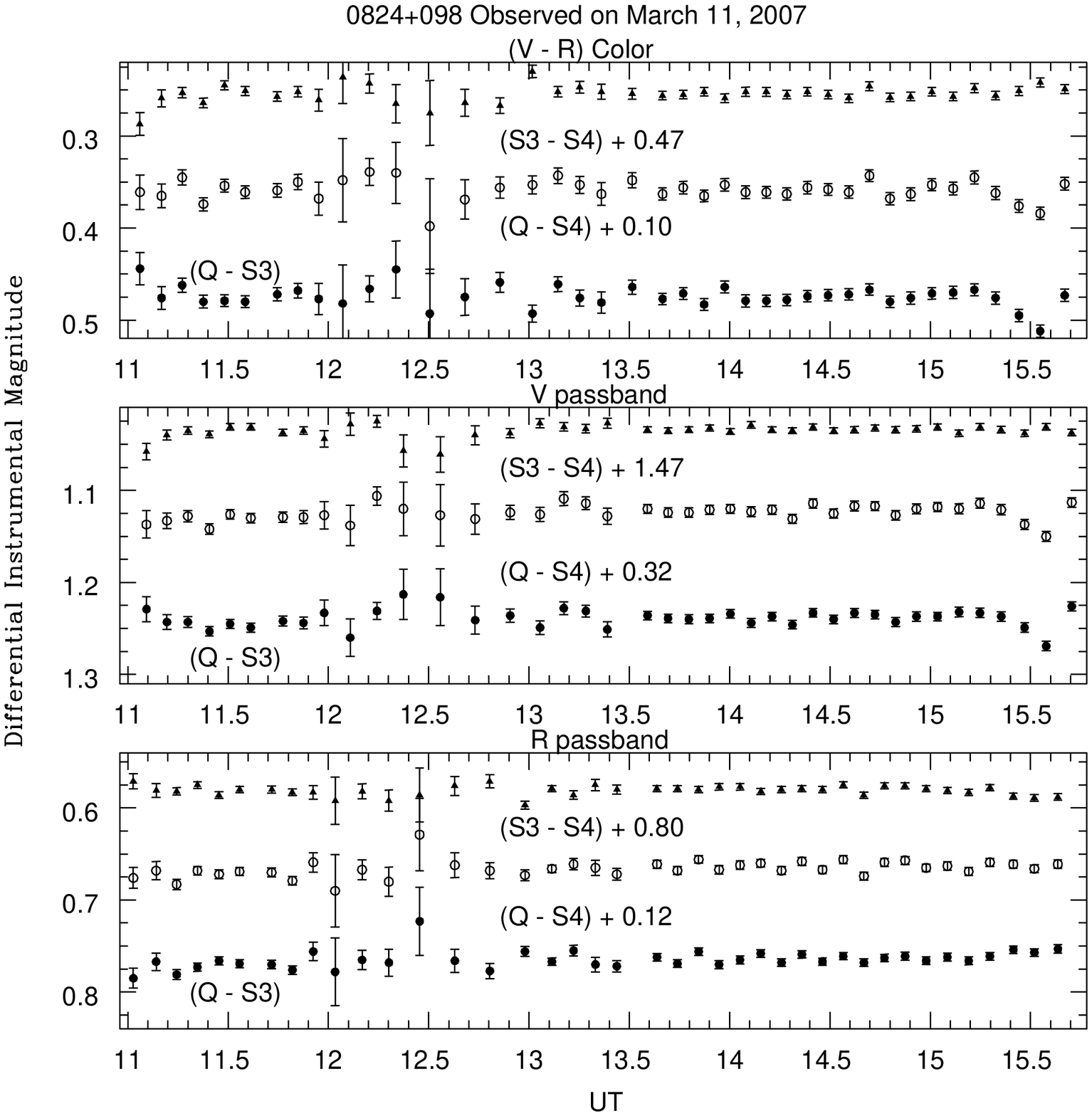}
\caption{The V, R and V$-$R light curves of 0824$+$098 on the night of March 11, 2007.
In caption S and Q represent star and QSO respectively. For clarity, the DLCs are offseted
by the amounts marked on the panels.}
\end{figure}

\begin{figure}
\plotone{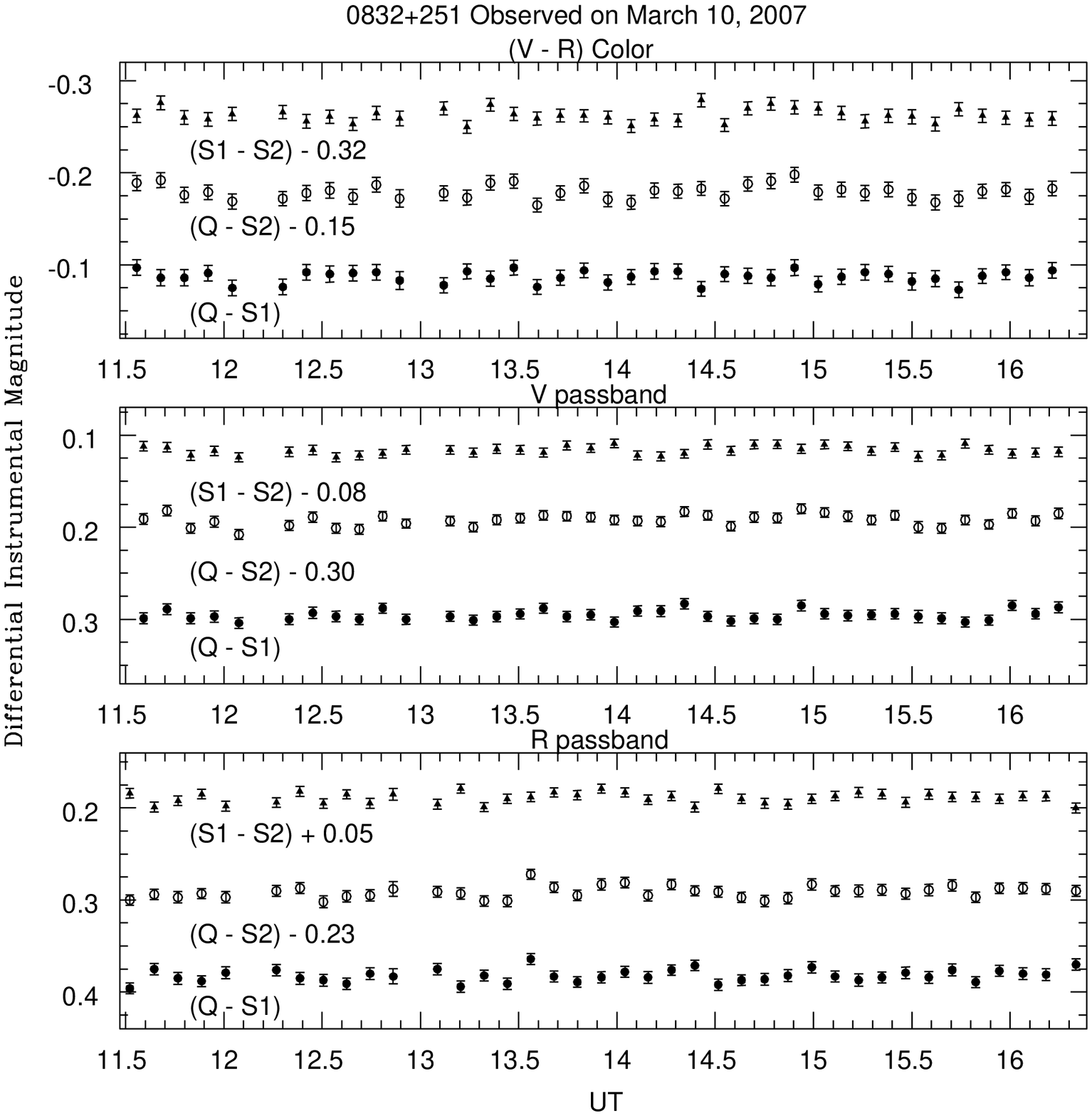}
\caption{The V, R and V$-$R light curves of 0832$+$251 on the night of March 10, 2007.
In caption S and Q represent star and QSO respectively. For clarity, the DLCs are offseted
by the amounts marked on the panels.}
\end{figure}

\begin{figure}
\plotone{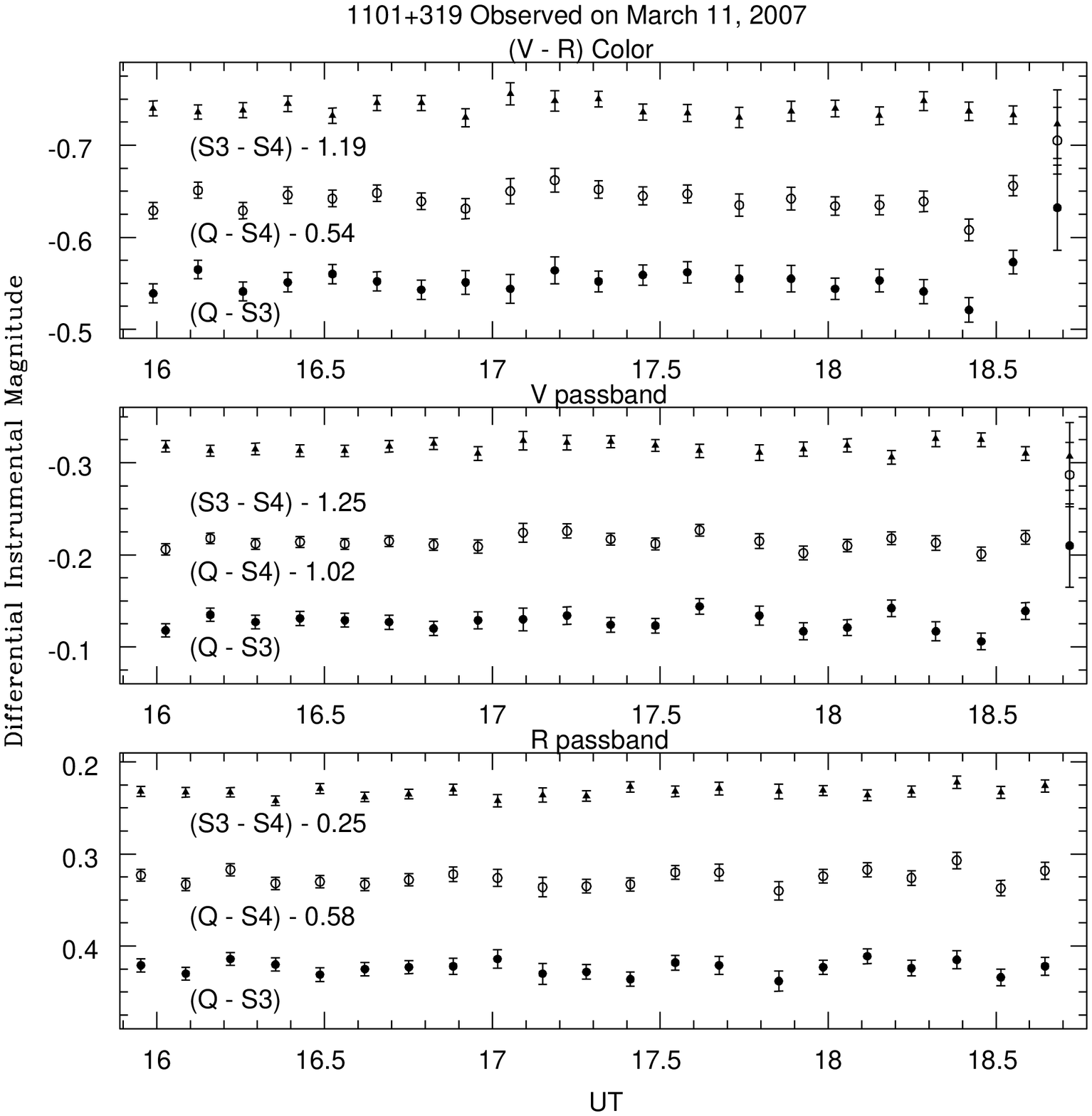}
\caption{The V, R and V$-$R light curves of 1101$+$319 on the night of March 11, 2007.
In caption S and Q represent star and QSO respectively. For clarity, the DLCs are offseted
by the amounts marked on the panels.}
\end{figure}

\begin{figure}
\plotone{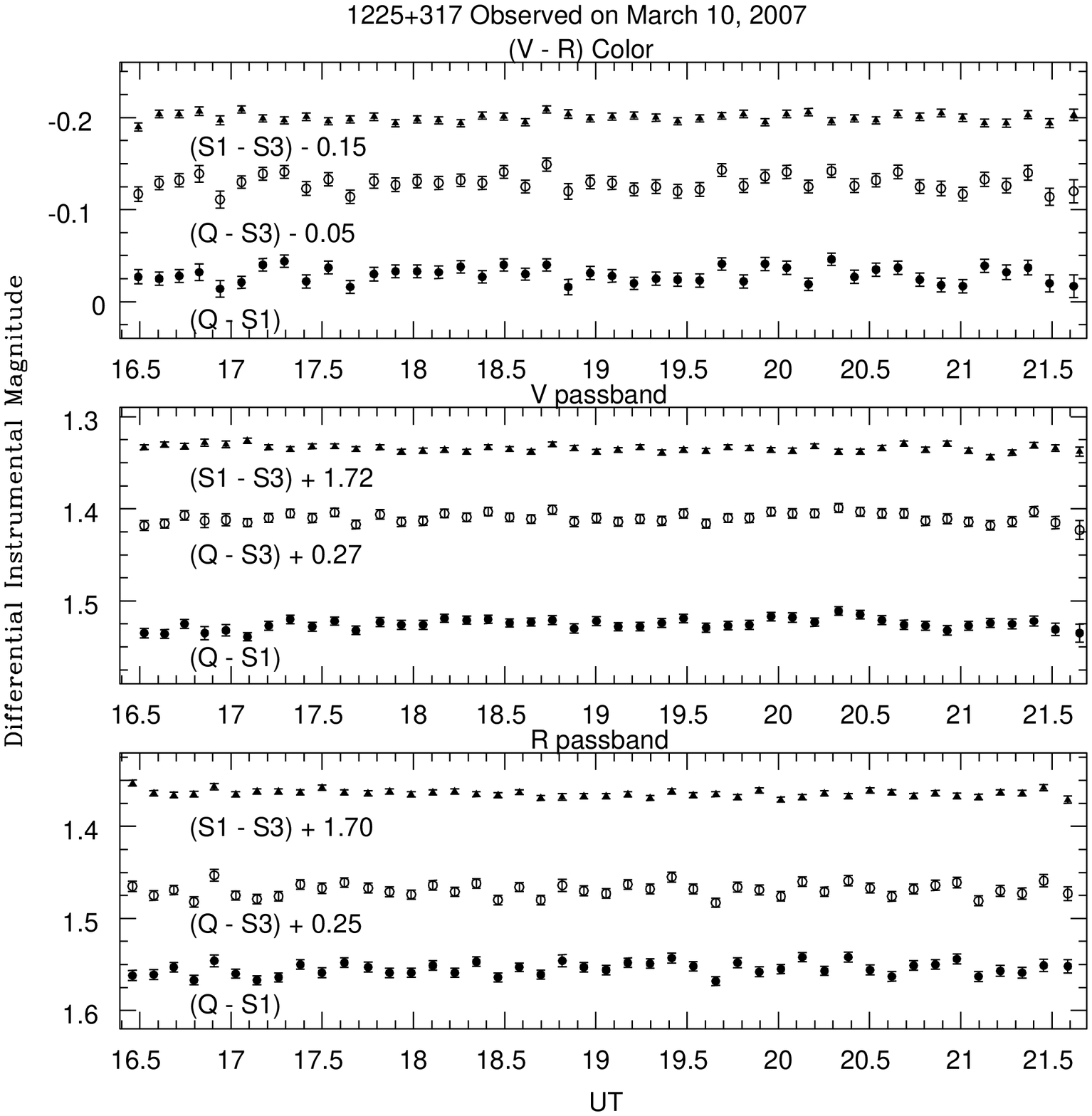}
\caption{The V, R and V$-$R light curves of 1225$+$317 on the nights of March 10, 2007.
In caption S and Q represent star and QSO respectively. For clarity, the DLCs are offseted
by the amounts marked on the panels.}
\end{figure}

\begin{figure}
\plotone{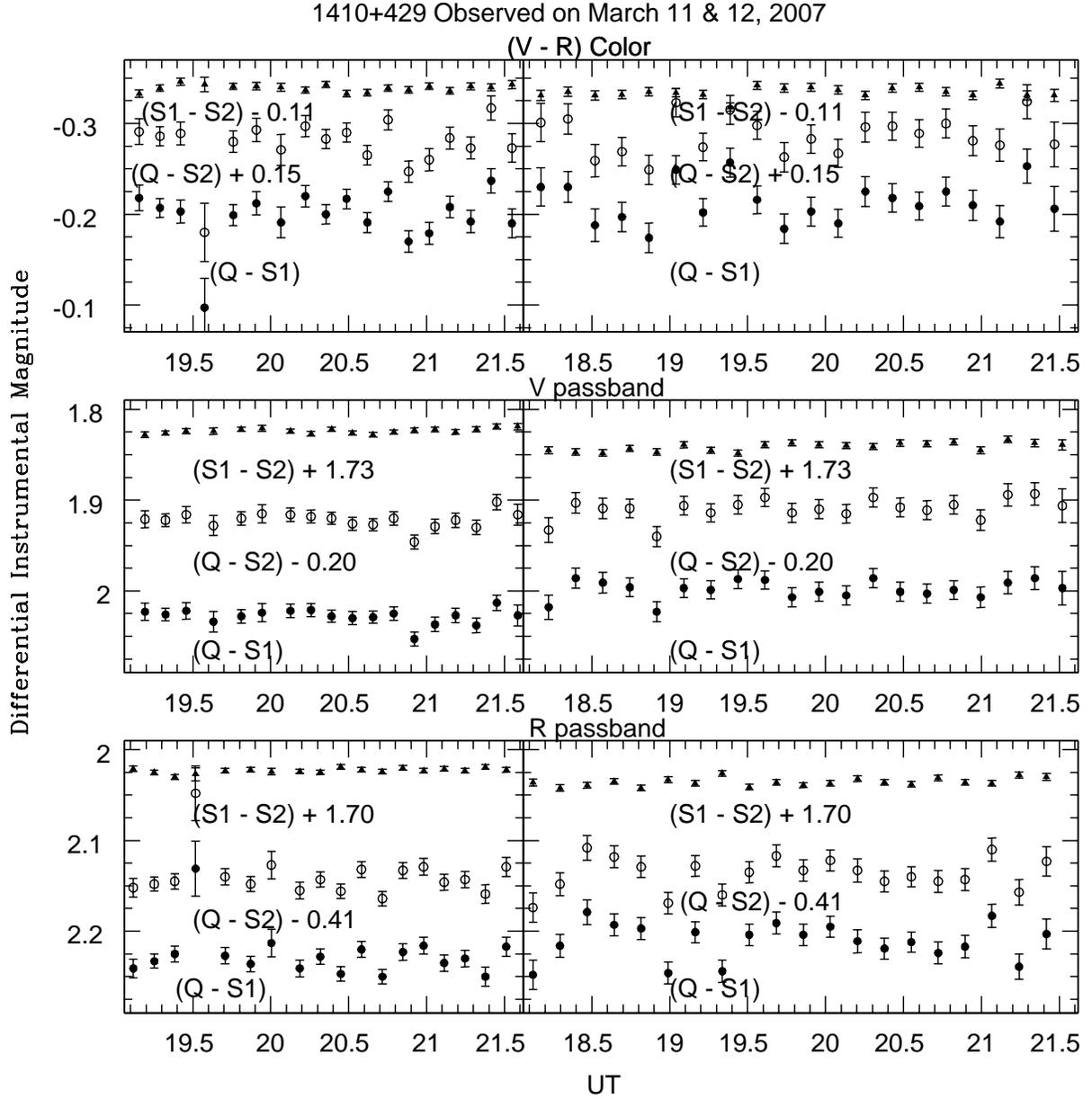}
\caption{The V, R and V$-$R light curves of 1410$+$429 on the nights of March 11 and 12, 2007.
In caption S and Q represent star and QSO respectively. For clarity, the DLCs are offseted
by the amounts marked on the panels. One point spike seen in R band QSO observations on March 11, 2007
which has caused one point dip in V$-$R color is omitted in variability detection test.}
\end{figure}

\end{document}